\documentclass[conference]{IEEEtran}
\usepackage{cite}
\usepackage{amsmath,amssymb,amsfonts}
\usepackage{algorithm}
\usepackage{graphicx}
\usepackage{textcomp}
\usepackage{color, xcolor}
\usepackage{algpseudocode}
\usepackage{subfigure}
\usepackage{stfloats}
\usepackage{float}

\newtheorem{remark}{Remark}
\def\BibTeX{{\rm B\kern-.05em{\sc i\kern-.025em b}\kern-.08em
    T\kern-.1667em\lower.7ex\hbox{E}\kern-.125emX}}
\begin{document}

\title{Dual Identities Enabled Low-Latency Visual Networking for UAV Emergency Communication}

\author{
	\IEEEauthorblockN{Yanpeng~Cui\textsuperscript{*}, Qixun~Zhang\textsuperscript{*}, Zhiyong~Feng\textsuperscript{*}, Zhiqing~Wei\textsuperscript{*}, Ce~Shi\textsuperscript{*}, Jinpo~Fan\textsuperscript{*} and Ping~Zhang\textsuperscript{*}}
	\IEEEauthorblockA{\textsuperscript{*}Key Laboratory of Universal Wireless Communications, Ministry of Education\\
		Beijing University of Posts and Telecommunications, Beijing, P.R.China, 100876.\\
		Email: \{cuiyanpeng94, zhangqixun, fengzy, weizhiqing, sc, fjp, pzhang\}@bupt.edu.cn}}

\maketitle

\begin{abstract}
	
The Unmanned Aerial Vehicle (UAV) swarm networks will play a crucial role in the B5G/6G network thanks to its appealing features, such as wide coverage and on-demand deployment. Emergency communication (EC) is essential to promptly inform UAVs of potential danger to avoid accidents, whereas the conventional communication-only feedback-based methods, which separate the digital and physical identities (DPI), bring intolerable latency and disturb the unintended receivers. In this paper, we present a novel DPI-Mapping solution to match the identities (IDs) of UAVs from dual domains for visual networking, which is the first solution that enables UAVs to communicate promptly with what they see without the tedious exchange of beacons. The IDs are distinguished dynamically by defining feature similarity, and the asymmetric IDs from different domains are matched via the proposed bio-inspired matching algorithm. We also consider Kalman filtering to combine the IDs and predict the states for accurate mapping. Experiment results show that the DPI-Mapping reduces individual inaccuracy of features and significantly outperforms the conventional broadcast-based and feedback-based methods in EC latency. Furthermore, it also reduces the disturbing messages without sacrificing the hit rate.
\end{abstract}


\section{Introduction}

To provide wide coverage and on-demand deployment services for B5G/6G mobile communication and wireless networking, the Unmanned Aerial Vehicle (UAV) network has been envisioned as a promising solution for public safety and disaster relief \cite{UAV_B5G}. However, UAVs are continuously exposed to events that might cause severe collisions \cite{UAV_Safety}. For instance, dozens of UAVs crashed into a building during a show in China's Chongqing city last year, and most of them might have been avoided if they had been notified of the potential disaster from nearby UAVs and responded promptly. Therefore, the need for communication-assisted collaborative perception naturally arises in UAV emergency communication (EC) \cite{ISAC}.

The EC between UAVs has been conventionally achieved via broadcasting. It generally causes confusion and disturbance at unintended receivers, as well as broadcast storm issues in large-scale networks\cite{Broadcast_Storm}. Therefore, unicast and multicast, which allow UAVs to communicate following a specific digital identity (D-ID), e.g. unique IP address, are crucial for EC. Note that it is not sufficient to only be aware of the D-ID, UAVs should have the capability to perceive the neighbor's physical identity (P-ID), e.g. location and velocity, since P-ID is more valuable than D-ID in EC. For instance, a UAV with obstacles ahead would like to alert the approaching ones behind, regardless of their IP addresses.

Building upon the communication-only feedback-based protocols, the conventional methods of P-ID acquisition require UAVs to periodically exchange beacons embedded with feature information \cite{Beacons}. However, it tends to be performed frequently in high-mobility scenarios to ensure sensing accuracy, which causes large EC latency and overhead. Developing a state prediction scheme can solve this issue to a certain extent, whereas the benefit depends on the accuracy of target's P-ID \cite{Prediction}. For instance, the positioning error may probably achieve several meters \cite{GNSS_Error}. Essentially, the above issue arises since the P-ID is passively ``heard" from the radio domain, and we thus vividly call it an auditory domain (AD) in the following. The sender should have the ability of active sensing to satisfy the low-latency requirement. At the time of writing, state-of-the-art techniques, such as circular scanning millimeter-wave (CSM) Radar, Lidar, laser range finder and zoom/wide camera, have been utilized on UAVs to sense the environment actively, we thus would say it is more like a visual domain (VD) relative to AD. Although it enhances the capability of recognition accuracy, perception range, spatial resolution, and robustness in extreme situations \cite{UAV_Sensors}, the difficulty in identifying D-ID is a common drawback of the VD-only sensing approaches. As a result, research efforts toward combining the sensing information from dual domains are well underway, and such examples can be found in \cite{EV_Loc} and \cite{Identity_and_WiFi_Mapping}. Although the sensing ability has been enriched, the digital and physical identities (DPI) are still separated thus far, which may cause intolerable delays and disturb the unintended receivers. 

To tackle the above problems, UAVs in EC are supposed to be endowed with an advanced capability: opening their eyes when communicating with neighbors by mapping DPI in AD and VD, which will bring the following benefits. First, having the combined IDs from dual domains may significantly reduce individual inaccuracy and unreliability while improving the overall performance. Moreover, it can provide the following service: given a specific P-ID, the D-ID of the matching UAV will be determined promptly and vice versa. Accordingly, alert messages can be accurately sent to UAVs of specific P-ID with the lowest latency and the least disruption to unintended UAVs. Nevertheless, it is not trivial to match what the UAVs hear and see due to the following challenges.

\begin{itemize} 
\item\textbf{Asynchrony}: Generally, the IDs observed from VD have a higher refresh rate than IDs obtained from AD, it thus may cause poor mapping performance since the matching of P-ID and D-ID cannot be performed at any time.
\item\textbf{Asymmetry}: Due to the different sensing results and range of AD and VD, UAVs would ``see" but not ``hear" others and vice versa, which causes matching difficulty.
\item\textbf{Similarity}: The features tend not to differ much sometime or somewhere, making it difficult to distinguish multiple UAVs. For instance, location is not an ideal P-ID if UAVs are close to each other or visually occluded.
\end{itemize}

In this paper, we present a novel DPI-Mapping approach, the first solution that enables UAVs to communicate instantly with what they see without the tedious exchange of beacons. To tackle the \textbf{Asymmetry} issue, we proposed a bio-inspired matching (BIM) algorithm to associate the ID pairs obtained from VD and AD and exploit them for prompt EC. We further consider a Kalman filtering (KF) scheme for tracking and predicting the state before obtaining the next D-ID to address the \textbf{Asynchrony} challenge. The P-IDs are distinguished via dynamic weight and associated accurately in different epochs to solve the \textbf{Similarity} problem. Experiment results show that our DPI-Mapping can remove the EC latency and minimize the disturbance rate without sacrificing the hit rate when compared to its broadcast and feedback-based counterparts.

\section{General Framework of DPI-Mapping}
\begin{figure}
	\centering
	{\includegraphics[width=0.90\columnwidth]{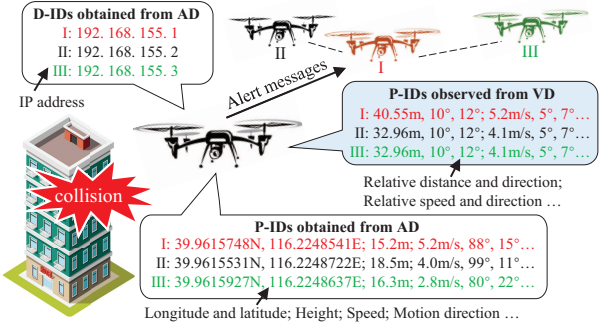}}
	\caption{The typical scenario for emergency communication in UAV networks.} \label{Scenario}
\end{figure}
\subsection{Motivation Scenario}
As shown in \textbf{Fig. \ref{Scenario}}, we consider a self-organized UAV swarm network where UAVs with obstacles ahead would like to alert the intended neighbors to avoid potential accidents. Each UAV measures several features constantly for itself and its neighbors. The unique IP address, longitude, latitude, height, speed and motion direction are embedded in beacons with a few bytes, which are broadcast periodically at an adaptive interval \cite{TARRAQ} to allow UAV to be discovered and recognized by neighbors from AD. Moreover, the relative distance, direction and relative velocity of neighbors can be observed from VD. Aiming at alerting intended neighbors promptly without disturbing the unintended receivers, UAVs should have the ability to communicate instantly with what they see without exchanging beacons tediously. This motivates us to present the following novel DPI-Mapping approach.

\subsection{Framework of the Proposed DPI-Mapping }
\begin{figure}
	\centering
	{\includegraphics[width=1\columnwidth]{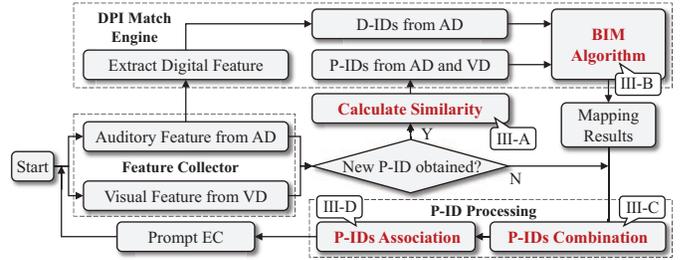}}
	\caption{The framework of DPI-Mapping. The critical steps are shown in red and will be discussed further in the section \textbf{\uppercase\expandafter{\romannumeral3}}.} \label{Framework}
\end{figure}

As shown in Fig. \ref{Framework}, the framework of our proposed DPI-Mapping is mainly composed of three modules: i) feature collector, ii) DPI match engine and iii) P-ID processing unit. 

\subsubsection{Feature collector}
The UAVs first receive the features from AD passively and detect the feature from VD actively via the feature collector. In addition to the D-IDs, DPI-Mapping also uses P-ID, namely a set of physical features to describe a UAV. Such features include wing type, body color, relative position and velocity, etc, whose reliability relies on their prevalence. For instance, the relative distance could distinguish UAVs in different positions while the relative velocity will be inoperative if they are relatively stationary.  Assuming that at the $n$th epoch, a UAV observes $K_v$ and $K_a$ neighbors' physical features from VD and AD, respectively, we denote $\textbf{VF}_n=\{\textbf{\textit{vf}}_n(i),i=1,...,K_v\}$ and $\textbf{AF}_n=\{\textbf{\textit{af}}_n(j),j=1,...,K_a\}$ as their P-IDs in VD and AD, where $\textbf{\textit{vf}}_n(i,k)$ and $\textbf{\textit{af}}_n(j,k)$ denote the $k$th physical feature ($k=1,...,K_f$) of $i$th and $j$th neighbor, respectively. 
Then they will be assigned with dynamic weights based on prevalence to calculate similarity, which will be discussed further in \textbf{\uppercase\expandafter{\romannumeral3}-A}.
\begin{remark}
The definition of P-ID is not limited to the mentioned example, any distinguishable feature is included in its connotation. The P-IDs could be observed from both VD and AD while the D-IDs will only be obtained from AD. 
\end{remark}

\subsubsection{DPI match engine}
\begin{figure}
	\centering
	{\includegraphics[width=0.9\columnwidth]{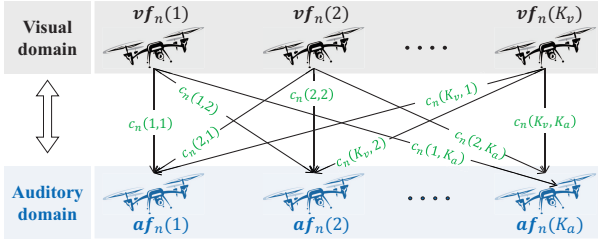}}
	\caption{Bipartite graph model for matching identity pairs in VD and AD.} \label{Bipartite graph model}
\end{figure}

The D-IDs together with the P-IDs will be fed into the bipartite graph matching model shown in Fig. \ref{Bipartite graph model}, where the edge's weight is defined as the matching cost $c_n(i,j)=s_n(i,j)^{-1}$, namely the reciprocal of similarity between $\textbf{\textit{vf}}_n(i)$ and $\textbf{\textit{af}}_n(j)$. The optimization target of P-ID pairs matching problem is given by $\min(f_1+f_2)$, where $f_1=\frac{1}{N}\sum\nolimits_{i=1}^{K_v}\sum\nolimits_{j=1}^{K_a}a_n(i,j)c_n(i,j)$ and $f_2=\frac{1}{N}\sqrt{\sum\nolimits_{i=1}^{K_v}\sum\nolimits_{j=1}^{K_a}\left(a_n(i,j)c_n(i,j)-f_1\right)^2}$ denote the sub-problems of minimizing the overall cost and equalizing the individual cost, respectively, where $N=\min(K_a,K_v)$. The constraints of $\sum\nolimits_{i=1}^{K_v}a_n(i,j)=1, j=1,...,K_a$ and $\sum\nolimits_{j=1}^{K_a}a_n(i,j)=1, i=1,...,K_v$ indicate that any P-ID in AD can only be matched with one in VD and vice versa. $\textbf{\textit{A}}_n$ is the assignment matrix, where $a_n(i,j)=1$ if $\textbf{\textit{vf}}_n(i)$ is matched with $\textbf{\textit{af}}_n(j)$, otherwise $a_n(i,j)=0$. Then the BIM algorithm (as presented in \textbf{\uppercase\expandafter{\romannumeral3}-B}), which matches neighbor's P-IDs in VD and AD according to similarity, is executed to solve the above optimization problem. 
\subsubsection{P-ID processing}
Based on the matching results of the BIM algorithm, D-IDs from AD are successfully mapped with P-IDs in VD since they have been naturally associated with P-IDs in AD. Then the mapped P-ID pairs from dual domains are combined via the KF method to obtain a more accurate result. The BIM algorithm will not be re-executed until new P-IDs arrive, and the KF will also be exploited to combine and predict P-IDs  (see more details in \textbf{\uppercase\expandafter{\romannumeral3}-C}). Finally, the P-IDs obtained from VD at different epochs are associated to avoid the mapping errors, which will be introduced in \textbf{\uppercase\expandafter{\romannumeral3}-D}.

\section{Matching, Prediction and Association of IDs in DPI-Mapping}
In order to describe the above key steps of DPI-Mapping in detail and tackle the \textbf{Asynchrony}, \textbf{Asymmetry} and \textbf{Similarity} issues, we present the corresponding approaches as follows.
\subsection{Distinguish Features via Dynamic Weight}
Given two P-IDs $\textit{\textbf{vf}}_n(i)$ and $\textit{\textbf{af}}_n(j)$, we want to calculate $s_n(i,j)$, namely the probability that they correspond to the same UAV in reality. However, it is particularly challenging due to the \textbf{Similarity} issue, namely their distinguishability generally depends on the dynamic environment. For instance, the difference in relative velocity is more distinctive if UAVs are flying freely, and the relative position has low importance in distinguishing when UAVs are close to each other. This motivates us to present a novel dynamic weight for various physical features based on their prevalence. 

We define a cosine similarity function $D\left(\textit{\textbf{f}}_n(a),\textit{\textbf{f}}_n(b)\right)=\textit{\textbf{f}}_n(a)\cdot\textit{\textbf{f}}_n(b)/\left\|\textit{\textbf{f}}_n(a)\right\|\left\|\textit{\textbf{f}}_n(b)\right\|$ to denote the similarity of two physical features $\textit{\textbf{f}}_n(a)$ and $\textit{\textbf{f}}_n(b)$. When they come from the same domain, it represents the similarity of two different UAVs. When they come from various domains, it describes how likely they are two measurements of the same UAV. Without loss of generality, we assume that $\textbf{\textit{vf}}_n(k)=\{\textbf{\textit{vf}}_n(i,k),i=1,...,K_v\}$ denotes a vector of all measurement of the $k$th physical feature on $K_v$ neighbors. 
The weight of $\textbf{\textit{vf}}_n(k)$ is assigned as $w_k=\frac{1}{K_v}\sum\nolimits_{i=1}^{K_v}p_n(i,k)$, where $p_n(i,k)=\sum\nolimits_{j\neq i}D(\textbf{\textit{vf}}_n(i,k), \textbf{\textit{vf}}_n(j,k))\prod\nolimits_{m\neq j,m\neq i}(1-D(\textbf{\textit{vf}}_n(i,k), \textbf{\textit{vf}}_n(m,k)))$
denotes the distinguishability of $\textbf{\textit{vf}}_n(i,k)$, namely the probability that $\textbf{\textit{vf}}_n(i,k)$ is different from other features in $\textbf{\textit{vf}}_n(k)$. Then the similarity of $\textit{\textbf{vf}}_n(i)$ and $\textit{\textbf{af}}_n(j)$ is defined as the harmonic mean of individuals 
\begin{equation}\label{Similarity} 
	s_n(i,j)=\left\{\sum\nolimits_{k=1}^{K_f}w_k^{\prime}/D\left(\textbf{\textit{vf}}_n(i,k), \textbf{\textit{af}}_n(j,k)\right)\right\}^{-1}, 
\end{equation}
where $w_k^{\prime}$ is the normalized weight. 

In this way, the features are distinguished dynamically via their prevalence. Moreover, if two UAVs have high dissimilarities in most features, the similarity will be low despite the large weights of others, and the outliers could also be mitigated.

\subsection{Match D-ID and P-ID Pairs by BIM Algorithm}

This subsection proposes a bio-inspired matching (BIM) algorithm to accurately match the P-IDs and D-IDs pairs. Vampire bats have various interests in different preys depending on tastes and hunting risk. They tend to compete for favorite prey when hunting in groups, whereas feeding the extra food back to companions in starvation according to kinship \cite{Vampire Bat Optimizer}. The above selfish and altruistic behavior actually achieves the maximization of group income and the balance of individual benefits, which is similar to our aim that achieving the overall minimization and the individual equalization of matching cost without restriction on the number equality of P-ID parties. 

For the task of mapping P-ID pairs from VD and AD, we regard $\textbf{VF}_n$ and $\textbf{AF}_n$ as vampire bats and preys, respectively, and present a specific mapping process as follows.

\subsubsection{Initially match P-IDs in VD and AD}

Note that a bat cannot catch multiple preys at a time, and a prey cannot be shared by multiple bats, despite the probably unequal number of bats and prey. Each bat thus evaluates the benefits of all accessible preys and selects the one with the highest profit as the initial competitive target before hunting.

Inspired by the bio-behavior, the cost of matching $\textbf{\textit{vf}}_n(i)\in\textbf{VF}_{n}$ and $\textbf{\textit{af}}_n(j)\in\textbf{AF}_{n}, \forall i,j$ will be calculated via $c_n(i,j)=s_n(i,j)^{-1}$, where $s_n(i,j)$ is the similarity defined in (\ref{Similarity}). Even if there are differences between measurements in AD and VD due to the \textbf{Asymmetry} issue, we believe that two P-IDs will be ``close to" each other if they belong to the same UAV. Thus the one with the lowest cost, namely the best matching of $\textbf{\textit{vf}}_n(i)$ is determined by
\begin{equation}\label{Initial matching} i=\mathop{\arg\min}_{j=1,...,K_a}\textit{c}_n(i,j)=\mathop{\arg\min}_{j=1,...,K_a}\textit{s}_n(i,j)^{-1}.
\end{equation}
\begin{remark}
Note that the \textbf{Asymmetry} issue will also arise when fewer P-IDs are observed in one domain (deficient domain $D_d$) than in the other (complete domain $D_c$). To equalize their numbers, the cost matrix is augmented by adding virtual P-IDs in $D_d$, whose matching costs to P-IDs in $D_c$ are set to infinity, thus the superfluous P-IDs in $D_c$ will not be mapped erroneously.
\end{remark}
\subsubsection{Competition}
There are always competitions between bats since their favorite prey often conflicts, and the expected benefit of their favorite prey will be decreased gradually during competition. A bat will find another better prey if the current benefit is no longer the best, and then continue to compete until all prey no longer conflict.

Similarly, the conflict also occurs once a P-ID (e.g. $\textbf{\textit{af}}_n(j)$) in a domain corresponds to the best matching of several P-IDs in the other domain (e.g., $\textbf{\textit{vf}}_n(i),i\in\Phi_n$, where $\Phi_n$ denotes the set of conflicting P-IDs), which is common after step 1). Inspiring by the competitive behavior of bats, we update the mapping cost to remove the conflict via
\begin{equation}\label{Competition}
	\begin{aligned}
		c^{t+1}_n(i,j)&\leftarrow c^{t}_n(i,j)\\&+\alpha\left[\min\limits_{i\in\Phi_n}c^t_n(i,j)-\min\limits_{i'\neq i,i'\in\Phi_n}c^t_n(i',j)+\epsilon^{t}\right],
	\end{aligned}
\end{equation}
where $\alpha$ is the competing rate. The first two items in parentheses denote the minimum and second minimum cost of matching $\textbf{\textit{af}}_n(j)$ from $\textbf{\textit{vf}}_n(i),i\in\Phi_n$, and $\epsilon$ is set to avoid the update failure due to their equality. The competition will not be terminated until the conflict no longer exists.

\subsubsection{Exchange the matching results}

Although the overall income of all bats is maximized after the competition, the individual benefits are not equal. Interestingly, a satiate bat will vomit excess blood to a proper companion based on kinship. 

Analogously, let's exchange the matching results after step 2) to balance the matching cost. For any two P-ID pairs that have been successfully matched, e.g. $\textbf{\textit{vf}}_n(i)$ and $\textbf{\textit{af}}_n(j)$, $\textbf{\textit{vf}}_n(p)$ and $\textbf{\textit{af}}_n(q)$, the assignment matrix will be updated by $a_n(i,j)=a_n(p,q)=0$ and $a_n(i,q)=a_n(p,j)=1$
if condition $\textit{c}_n(i,j)\geq\max\left\{\textit{c}_n(i,q),\textit{c}_n(p,j)\right\}$ is met. We collect the matching pairs that satisfy the above condition in a set $\Psi_{i,j}$ and find the best one with the minimum lost by
\begin{equation}\label{Multi_Exchange} j=\mathop{\arg\min}_{m\in\Psi_{i,j}}\left(\max\left\{c_n(i,q_m),c_n(p_m,j)\right\}\right).
\end{equation}
In this way, the assignment matrix namely the best mapping results of P-ID pairs can be determined. The pseudo-code of the BIM is presented in \textbf{Algorithm} \ref{Pseudo Code}. 

Finally, based on the above matching results and the natural association between D-IDs and P-IDs in AD, the D-IDs from AD are successfully mapped with P-IDs in VD. 
\begin{algorithm}[h]
	\caption{BIM algorithm}  
	\label{Pseudo Code}
	\begin{algorithmic}[1]
		\Require P-IDs from AD and VD
		\Ensure Assignment matrix $\textbf{\textit{A}}_n$
		\While{mapping conflicts exist}
		\For{$i$ from $1$ to $\max\{K_v,K_a\}$}
		\State Find the best matching results of $\textbf{\textit{vf}}_{i,n}$ via (\ref{Initial matching})
		\EndFor
		\For{all P-IDs in $\Phi_n$}
		\State Update the mapping cost based on (\ref{Competition})
		\EndFor
		\EndWhile
		\While{the exchange condition is met}
		\For{all P-IDs in $\Psi_n$}
		\State Find the optimum pair via (\ref{Multi_Exchange})
		\State Update the assignment matrix
		\EndFor
		\EndWhile
	\end{algorithmic}
\end{algorithm}
\subsection{Combination and Prediction of P-ID for a Single UAV}

Recall in \textbf{\uppercase\expandafter{\romannumeral2}-B}, we usually only have P-IDs from VD owing to the \textbf{Asynchrony} issue, thus the BIM algorithm will not be re-executed until a new P-ID is obtained from AD. The matched P-IDs should be combined after BIM to obtain more accurate results and then predicted to ensure the accuracy until the next BIM process arrives. To this end, we consider a KF approach for P-ID combination and estimation. 

Assuming a constant velocity mobility model, the evolution and measurement of UAV's state are given by $\boldsymbol{x}_n=\textbf{F}\boldsymbol{x}_{n-1}+\boldsymbol{u}_{n-1}$ and $\boldsymbol{y}_n=\textbf{H}\boldsymbol{x}_{n-1}+\boldsymbol{z}_{n-1}$, where  $\boldsymbol{x}_{n}=[p_n(1),p_n(2),p_n(3),v_n(1),v_n(2),v_n(3)]^T$ denotes the position and speed. $\boldsymbol{x}_{n-1}$ denotes the state information embedded in beacons or predicted at the last epoch. $\boldsymbol{u}$ and $\boldsymbol{z}$ are zero-mean Gaussian distributed noise. The measurement of P-IDs, such as distance $r$, azimuth angle $\theta$ and elevation angle $\varphi$, is generally performed in a polar coordinate system while the state evolution model is generally established in a Cartesian one. Therefore, before using the standard KF procedure, the measurement information should be converted via $p_n(1)=r\hbox{cos}\varphi\hbox{cos}\theta$, $p_n(2)=r\hbox{cos}\varphi\hbox{cos}\theta $ and $p_n(3)=r\hbox{sin}\varphi$. 

Considering that the conversion error will lead to a biased estimation, which will degrade filtering performance \cite{MUCM}. We thus propose a modified unbiased converted measurements KF (MUCM-KF) approach that enables unbiased conversion. The modified unbiased conversion for measurement can be given by $p_n(1)=\lambda_\varphi^{-1}\lambda_\theta^{-1}r\hbox{cos}\varphi\hbox{cos}\theta$, $p_n(2)=\lambda_\varphi^{-1}\lambda_\theta^{-1}r\hbox{cos}\varphi\hbox{sin}\theta$ and $p_n(3)=\lambda_\theta^{-1}r\hbox{sin}\varphi$. Then the measurement-conditioned mean of the converted measurement error is given by $\boldsymbol{\mu} =E\left[\textbf{p}_n|r,\theta,\varphi\right]	=	[(\lambda_{\theta}^{-1}\lambda_{\varphi}^{-1}-\lambda_{\theta}\lambda_{\varphi})r{\hbox{cos}}\theta{\hbox{cos}}\varphi, (\lambda_{\theta}^{-1}\lambda_{{\varphi}}^{-1}-\lambda_{\theta}\lambda_{{\varphi}})r{\hbox{sin}}\theta{\hbox{cos}}\varphi, (\lambda_{{\varphi}}^{-1}-\lambda_{{\varphi}})r{\hbox{sin}}\varphi]^T$, and the symmetric covariance matrix $\textbf{R}_{3\times3}$ is given by
\begin{equation}\label{MUCMKF}
	\left\{
	\begin{aligned}
		R(1,1)=\ &(r^{2}+\sigma_{r}^{2})(1+\lambda_{\theta}^{\prime}{\hbox{cos}} 2\theta)(1+\lambda_{{\varphi}}^{\prime}{\hbox{cos}} 2{\varphi})/4\\
		&-\lambda_{\theta}^{2}\lambda_{{\varphi}}^{2}r^{2}{\hbox{cos}}^{2}\theta{\hbox{cos}}^{2}\varphi\\
		R(1,2)=\ &(r^{2}+\sigma_{r}^{2})\lambda_{\theta}^{\prime}{\hbox{sin}} 2\theta(1+\lambda_{{\varphi}}^{\prime}{\hbox{cos}} 2{\varphi})/4\\  
		&-\lambda_{\theta}^{2}\lambda_{{\varphi}}^{2}r^{2}{\hbox{sin}}\theta{\hbox{cos}}\theta{\hbox{cos}}^{2}{\varphi}\\
		R(1,3)=\ &(r^{2}+\sigma_{r}^{2})\lambda_{\theta}\lambda_{{\varphi}}^{\prime}{\hbox{cos}}\theta{\hbox{sin}} 2{\varphi}/2\\ 
		&-\lambda_{\theta}\lambda_{{\varphi}}^{2}r^{2}{\hbox{cos}}\theta{\hbox{sin}}{\varphi}{\hbox{cos}}{\varphi}\\
		R(2,2)=\ &(r^{2}+\sigma_{r}^{2})(1+\lambda_{\theta}^{\prime}{\hbox{cos}} 2\theta)(1+\lambda_{{\varphi}}^{\prime}{\hbox{cos}} 2{\varphi})/4\\ 
		&-\lambda_{\theta}^{2}\lambda_{{\varphi}}^{2}r^{2}{\hbox{sin}}^{2}\theta{\hbox{cos}}^{2}\varphi\\
		R(2,3)=\ &(r^{2}+\sigma_{r}^{2})\lambda_{\theta}\lambda_{{\varphi}}^{\prime}{\hbox{sin}}\theta{\hbox{sin}} 2{\varphi}/2\\ 
		&-\lambda_{\theta}\lambda_{{\varphi}}^{2}r^{2}{\hbox{sin}}\theta{\hbox{sin}}{\varphi}{\hbox{cos}}{\varphi}\\
		R(3,3)=\ &(r^{2}+\sigma_{r}^{2})(1-\lambda_{{\varphi}}^{\prime}{\hbox{cos}} 2{\varphi})/2-\lambda_{{\varphi}}^{2}r^{2}{\hbox{sin}}^{2}{\varphi}
	\end{aligned}
	\right.,
\end{equation}
where $\lambda_x=e^{-\sigma_{x}^2/2}$, $\lambda_x^\prime=e^{-2\sigma_{x}^2}$, and $\sigma_{x}$ denotes the standard deviation of $x$.

The proposed MUCM-KF is designed by applying the unbiased conversion and covariance matrix to the standard KF for feature estimation, whose specific steps will not be introduced here due to page limitations, and readers can refer to our recent work \cite{TARRAQ} for more details.

\subsection{P-ID Association for Multiple UAVs}
Another critical issue raised in the multi-UAV scenario is to distinguish multiple neighbors accurately to realize the following two aims: i) to send the correct EC message to the intended UAVs and ii) to correctly match the measurement to the corresponding predicted state. Note that the mapping results might be disrupted later due to the \textbf{Similarity} issue, we thus propose a simple P-ID association approach as follows.

At the $n$th epoch, the UAV detects $K_v$ neighbors from VD via feature collector, which are processed to formulate $K_v$ measurements $\boldsymbol{y}_n(i), i=1,...,K_v$. The estimate $\hat{\boldsymbol{x}}_n(i)$ would be no longer accurate once the $\boldsymbol{y}_n(i)$ is wrongly associated with a state prediction $\hat{\boldsymbol{x}}_{n|n-1}(j),j\neq i$. Accordingly, the prediction at the ($n+1$)th epoch would be erroneous. Note that $\boldsymbol{y}_n(i)$ and $\hat{\boldsymbol{x}}_{n|n-1}(i)$ will not be ``far from" each other, we first calculate all the estimation of measurement of $K_v$ state predictions by $\hat{\boldsymbol{y}}_{n|n-1}(j)=\textbf{H}\hat{\boldsymbol{x}}_{n|n-1}(j),j=1,...,K_v$. For the measurement $\boldsymbol{y}_n(k)$ of $k$th neighbor, the difference between $\boldsymbol{y}_n(k)$ and each $\hat{\boldsymbol{y}}_{n|n-1}(j)$ is calculated by associating them via the smallest cosine distance principle
\begin{equation}\label{Association}
		i=\mathop{\arg\min}\limits_{j}D\left(\boldsymbol{y}_n(k),\textbf{H}\hat{\boldsymbol{x}}_{n|n-1}(j)\right),
\end{equation}
where the cosine similarity function is defined in \textbf{\uppercase\expandafter{\romannumeral3}-A}.

In this way, the state predictions at $n$th and $(n+1)$th epochs are correctly associated. $\hat{\boldsymbol{x}}_{n|n-1}(i)$ will be updated according to the correct measurement with the maximum accuracy while the D-IDs can be mapped to the correct P-IDs accordingly, which achieves the above-mentioned aims.

\section{Experiment and Simulation}
\subsection{Real-World Experiments}
\begin{figure}
	\centering
	\subfigure[DJI M300 RTK]{\includegraphics[width=0.48\columnwidth]{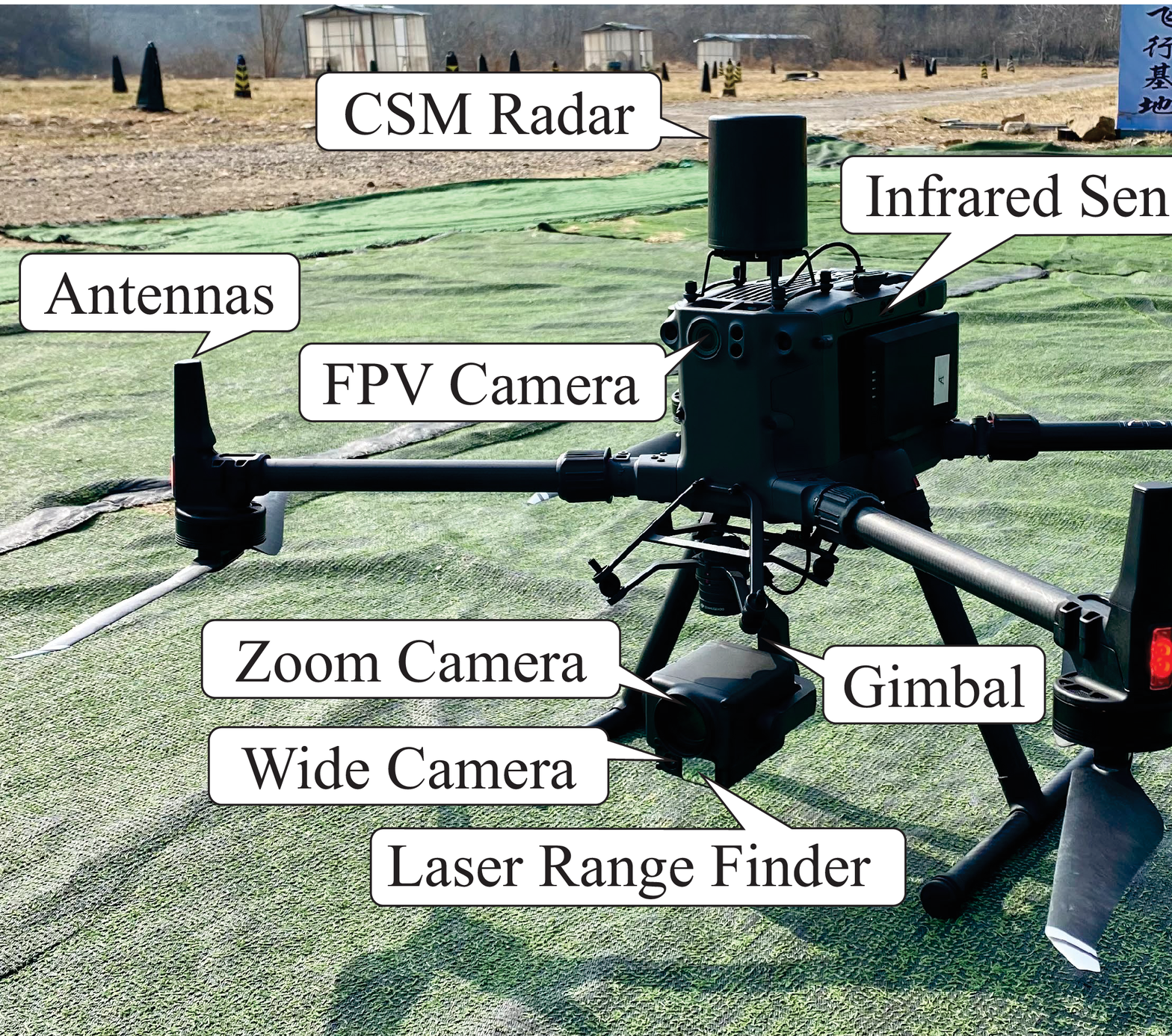}}
	\subfigure[Detection result of ZF-F1200]{\includegraphics[width=0.48\columnwidth]{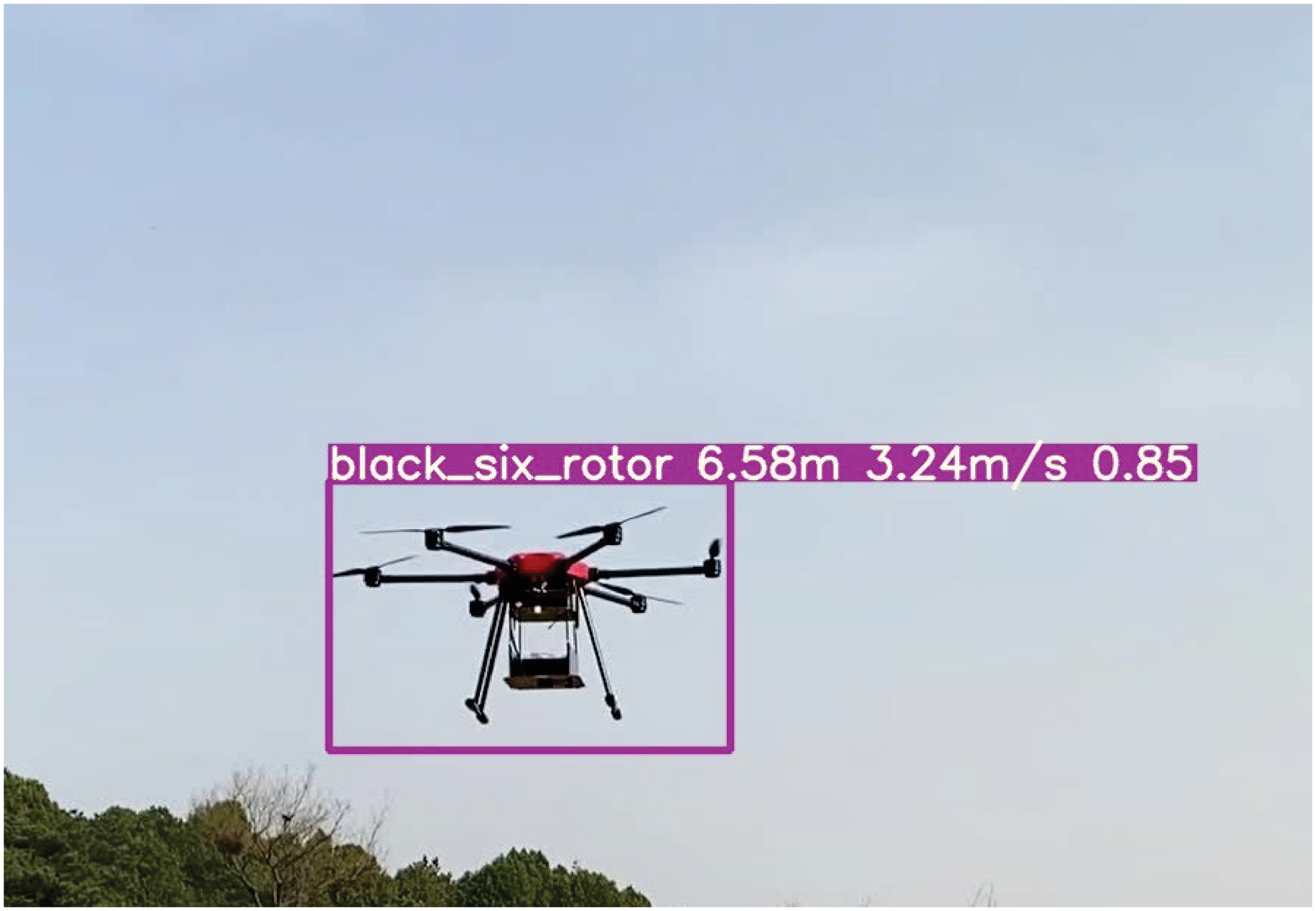}}
	\caption{UAV-based testbed for real-world experiments. } \label{M300}
\end{figure}
\begin{figure}[t]
	\centering
	\subfigure[RSSI versus distance]{\includegraphics[width=0.49\columnwidth]{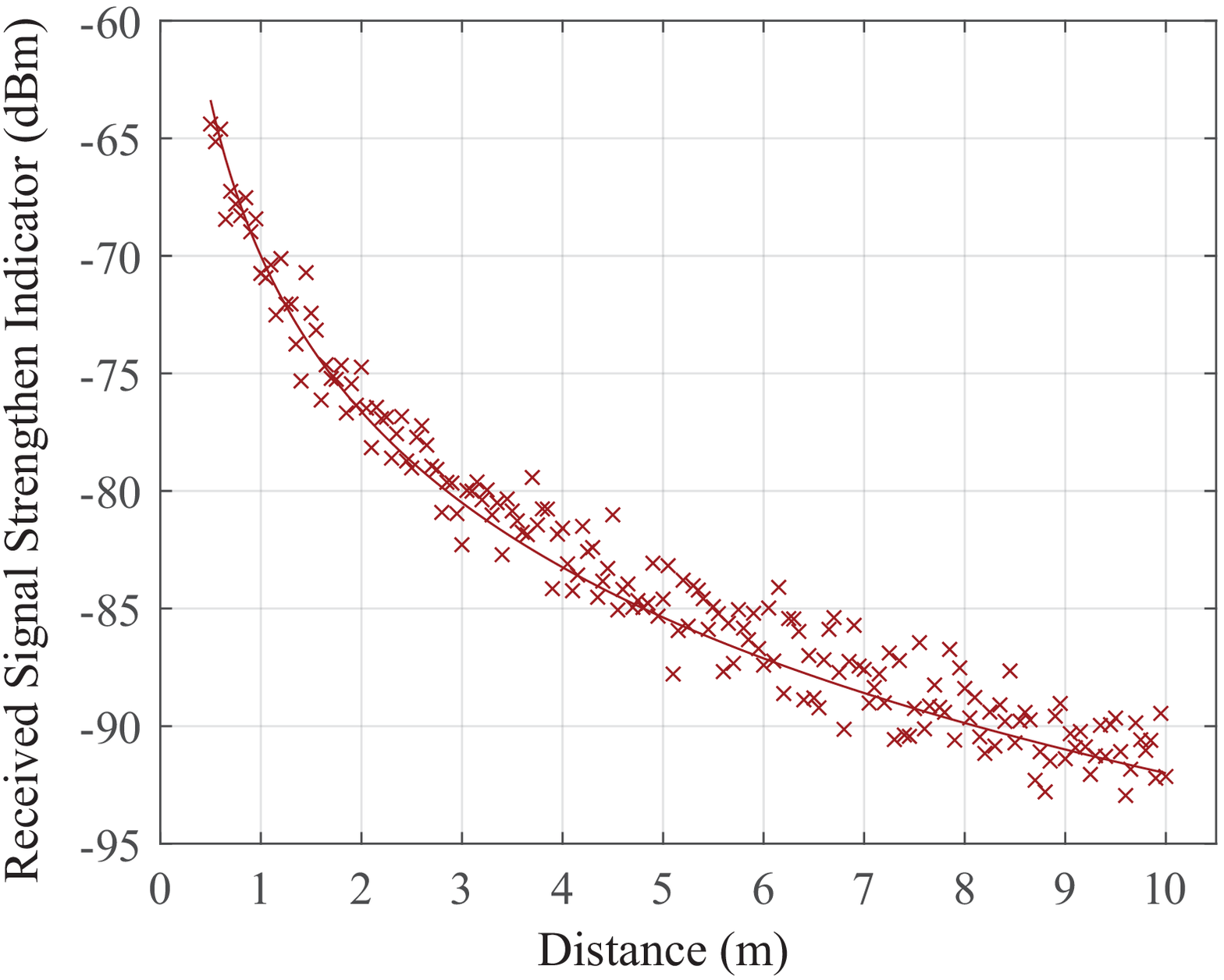}}
	\subfigure[Ranging error]{\includegraphics[width=0.49\columnwidth]{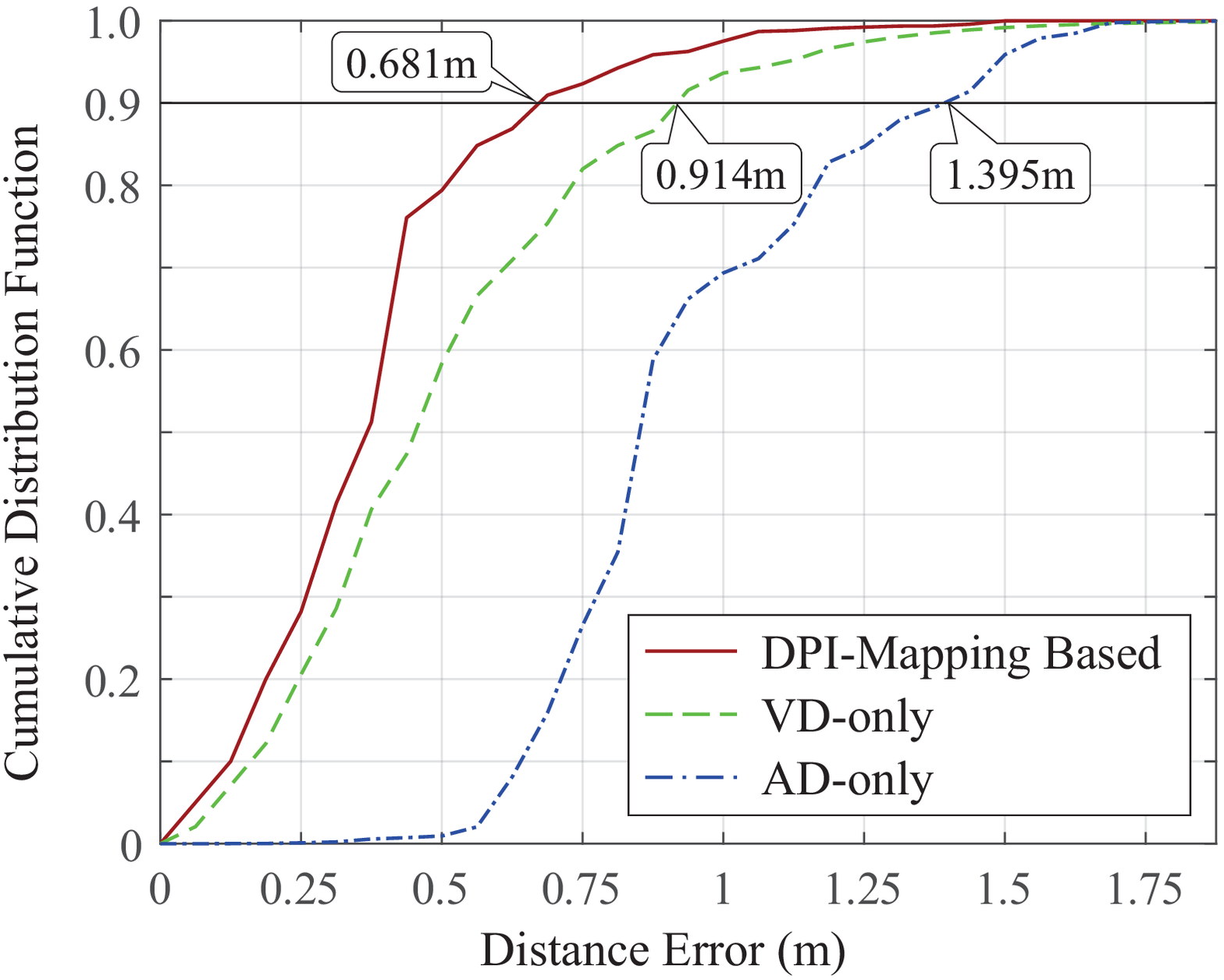}}
	\caption{Real-world experiment results of RSSI and ranging error.}\label{Real Experiment}
\end{figure}

For the real-world experiments, a DJI Matrice 300 (M300) RTK carrying CSM Radar, infrared sensing system and Zenmuse H20 gimbal \& camera system is used to detect potential dangers and alert neighbors, as shown in Fig. \ref{M300}(a). We applied You Only Look Once version 5 (YOLOv5) to detect ZF-F1200 (a six-rotor UAV with black color as shown in Fig. \ref{M300}(b)) from the video captured by M300 to build a VD-only sensing method. To realize the AD-only method, an Ettus USRP B210 connected to a mini-computer with an Intel Core i7-1165G7 CPU with OpenAirInterface platform was deployed in the pod of ZF-F1200. A Samsung Galaxy S6 is fixed in the pod of M300 to measure the Received Signal Strengthen Indicator (RSSI) via Cellular-Z. As shown in Fig. \ref{Real Experiment}(a), we've performed a lot of tests to establish the relationship between RSSI and distance in our target environments, which provides a basis for the RSSI-based AD-only ranging method.

By showing the cumulative distribution functions (CDF) of the average error of ranging, Fig. \ref{Real Experiment}(b) presents how DPI-Mapping improves the sensing accuracy. During the evaluation, M300 and ZF-F1200 hover at a height of 15m and 10m respectively, and the latter is 8m horizontally behind the former. Then we let them move towards the positive direction of M300 at 2m/s and 2.5m/s and ascend vertically at 2m/s and 2.5m/s respectively within 10s. The above trajectory is repeated 10 times with different initial locations. The proposed DPI-Mapping performs much better than AD-only and VD-only ranging methods since it combines P-IDs from dual domains and predicts the state accurately via MUCF-KF. The inferior performance of the VD-only method comes from the coordinate system conversion error and the accuracy of camera calibration. The worse results occur in the AD-only method due to the error of RSSI-based ranging. More specifically, with the indication of a laser range finder as the reference value, a 90 percentile errors of 0.68m are achieved by DPI-Mapping, which are lower than AD-only and VD-only ranging methods by 51.18\% and 25.49\%, respectively.

\subsection{Large-Scale Simulations}
\begin{figure}[t]
	\centering
	\subfigure[EC Latency]{\includegraphics[width=0.49\columnwidth]{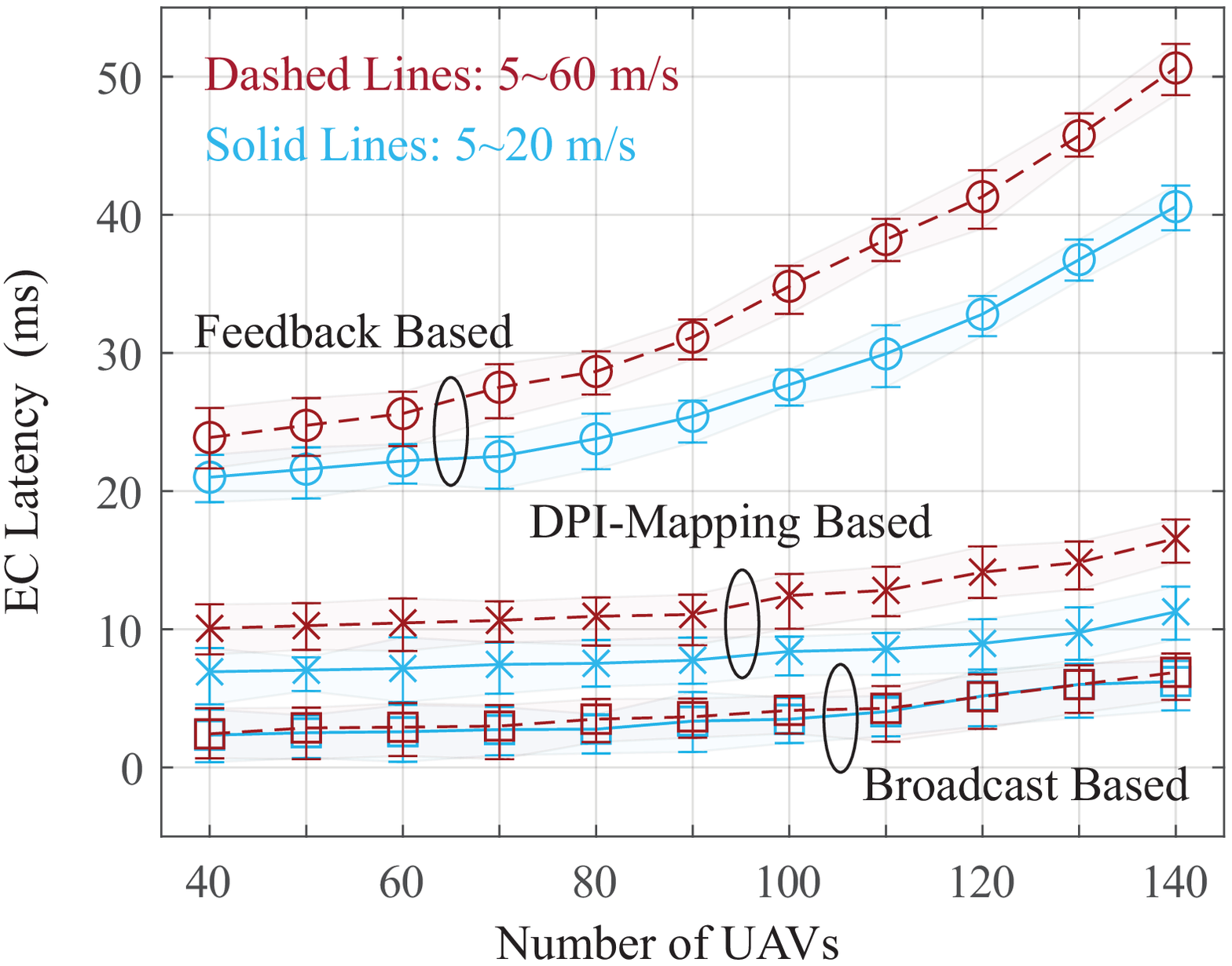}}
	\subfigure[Disturbance rate and hit rate]{\includegraphics[width=0.49\columnwidth]{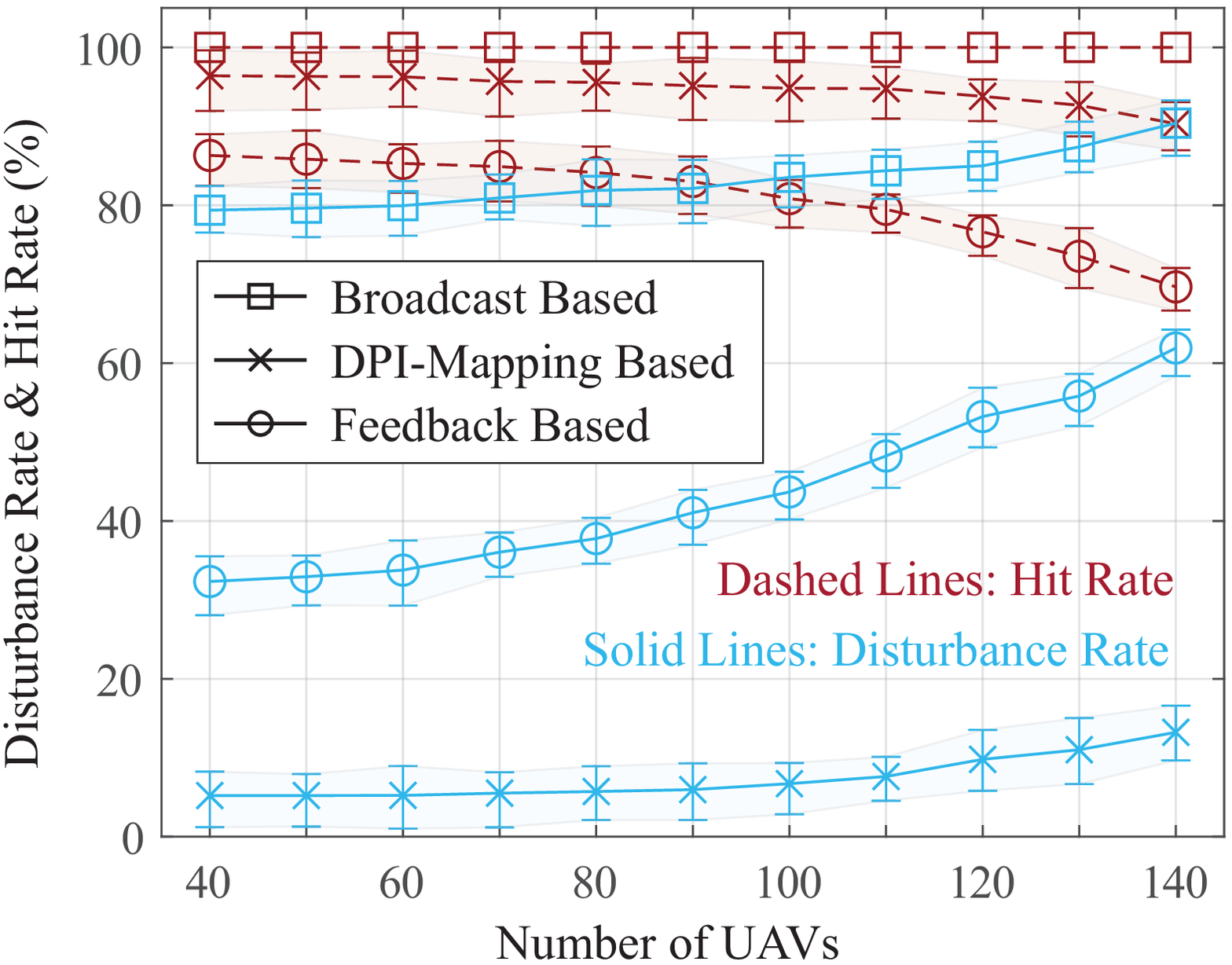}}
	\caption{Comparison results of the proposed DPI-Mapping and feedback and broadcast based methods in EC latency, disturbance rate and hit rate.}  \label{Simulation}
\end{figure}

It is difficult to verify EC latency and disturbance rate in the real world since we currently only have two UAVs. Therefore, we conduct large-scale simulations according to the simulation environment in \cite{TARRAQ}, namely 40$\sim$140 UAVs are moving with the random waypoint model in a 600m$\times$600m$\times$150m region. The lowest speed is 5m/s while the highest speed varies from 20m/s to 60m/s. EC messages are randomly generated on 10 UAVs per second, and the approaching neighbors behind are denoted as intended receivers. The AD-based ranging is achieved based on RSSI, whose standard deviation is set as $5/\sqrt{10}$ \cite{EV_Loc}. The visual appearances are simulated by image samples from datasets in \cite{{Data_Set}} and we scale them according to the relative distance \cite{EV_Loc}. We run 200 simulations and the results with 90\% confidence interval are presented in Fig. \ref{Simulation}. $\alpha=1$ and $\epsilon=0.02$ are set in BIM.

Fig. \ref{Simulation}(a) illustrates the EC latency, namely the time duration from the generation of the EC message to the successful reception by the intended UAV. There will be a worse latency in higher mobility since more beacons will be exchanged when the maximum speed increases. The broadcast has a superior latency of under 10ms since messages are sent to all nearby nodes without judging whether they are intended. Thanks to the accurate matching results of DPI-Mapping,  UAVs could communicate instantly with what they ``see" without the tedious exchange of beacons. The feedback-based approach requires UAVs to interact with neighbors before sending EC messages to confirm the D-ID, which inevitably causes extra latency. Although the EC latency rises in denser networks due to the increased exchanging of beacons and queuing latency, DPI-Mapping could bring down the latency to under 20ms since beacons are no longer required once the mapping is completed. As a result, the EC latency is reduced by 66.54\% on average when compared to the feedback-based method. Although the latency of DPI-Mapping is slightly worse than broadcast, the cost is still acceptable since it remarkably reduces the unnecessary disturbances, as described below.

For all the EC messages that have been sent, we denote the number of messages received by the intended and unintended UAVs as ${\rm RI}$ and ${\rm RU}$, and the number of EC messages not received by them as ${\rm NI}$ and ${\rm NU}$. ${\rm H}_r={\rm RI}/({\rm RI}+{\rm NI})$ and ${\rm D}_r={\rm RU}/({\rm RU}+{\rm RI})$ denote the hit rate and disturbance rate of EC messages, respectively. 

Fig. \ref{Simulation}(b) compares DPI-Mapping, the broadcast-based and the feedback-based approaches in terms of the performance of ${\rm H}_r$ and ${\rm D}_r$. As expected, all techniques show a better performance in the low-density network since high density means a high probability of confusing the intended UAVs with the unintended ones. We find that both the DPI-Mapping and the broadcast-based scheme are likely to hit a similar number of intended UAVs while the feedback-based one has a poor ${\rm H}_r$ since it generally misjudges the intended UAVs due to severe sensing error. It is worth highlighting that in our RSSI error model with $5/\sqrt{10}$ standard deviation, the feedback-based method will disturb a large number of unintended UAVs while the broadcast one enables all nearby UAVs within the communication range to receive the EC messages. DPI-Mapping is considerably better than them since it provides an interesting service, namely accurately confirming the D-ID according to the intended P-ID. It minimizes ${\rm D}_r$ without sacrificing ${\rm H}_r$ and thus outperforms the feedback-based approach by 36.29\% and 16.14\% in ${\rm D}_r$ and ${\rm H}_r$, respectively, and reduces ${\rm D}_r$ by 75.72\% when compared with the broadcast-based method.
\section{Conclusion}
In this paper, we have presented how to enable UAVs to communicate instantly with what they see in EC. The proposed DPI-Mapping approach distinguishes features by dynamic weight, maps D-IDs and P-IDs via the BIM algorithm, and combines and tracks the feature by the MUCM-KF. Experiment results show that it yields better sensing precision and lower latency, and significantly outperforms the broadcast and feedback based schemes in reducing disturbing messages without sacrificing the hit rate. Given the page limit, we designate multi-features-based P-ID definition, complexity and extra energy/computing consumption as our future work.
\section*{Acknowledgment}

This work was partly supported by National Key Research and Development Project (2020YFA0711300), National Natural Science Foundation of China (62022020, 61790553), and BUPT Excellent Ph.D. Students Foundation (CX2022208).

\vspace{12pt}

\end{document}